\documentclass[conference,a4paper,10pt]{IEEEtran}
\usepackage[top=1.8cm,bottom=1.8cm,left=1.5cm,right=1.5cm]{geometry}
\usepackage[utf8]{inputenc} 
\usepackage[T1]{fontenc}
\usepackage{url}
\usepackage{ifthen}
\usepackage{titlesec}
\usepackage[noadjust]{cite}
\usepackage[cmex10]{amsmath} 
\usepackage{amssymb}
\usepackage{graphicx}
\usepackage{mathtools}
\DeclarePairedDelimiter\ceil{\lceil}{\rceil}
\DeclarePairedDelimiter\floor{\lfloor}{\rfloor}
\usepackage{multirow}
\usepackage{array}
\usepackage{makecell}
\usepackage{tikz}
\usepackage{diagbox}
\usepackage{adjustbox}
\usepackage{caption, multirow, makecell}
\usepackage{algorithm,algorithmic}
\usepackage{amsthm}
\usepackage{comment}
\usepackage{subcaption}

\usepackage{tabularx}

\usepackage[english]{babel}
\newtheorem{thm}{Theorem}
\newtheorem{lem}{Lemma}

\newtheorem{defn}{Definition}

\newtheorem{exmp}{Example}

\setcellgapes{3pt}

\begin{document}
\title{An Improved Lower Bound for \\Multi-Access Coded Caching } 
	
\author{%
		\IEEEauthorblockN{K. K. Krishnan Namboodiri and B. Sundar Rajan}
		\IEEEauthorblockA{Department of Electrical Communication Engineering,\\ Indian Institute of Science, Bengaluru 560012, KA, India \\
			E-mail: \{krishnank, bsrajan\}@iisc.ac.in}
}
\maketitle
%%%%%%%%%%%%	

\begin{abstract}
	The multi-access variant of the coded caching problem with $N$ files, $K$ users and $K$ caches, where each user has access to $L$ neighbouring caches in a cyclic wrap-around manner, is considered. A cut-set based lower bound on the optimal rate-memory trade-off of the multi-access coded caching (MACC) scheme is derived. Furthermore, an improved lower bound on the optimal rate-memory trade-off of the MACC scheme is derived using non-cut-set arguments. The improved lower bound is tighter than the previously known lower bounds for the same setting.
\end{abstract}

%%\begin{IEEEkeywords}
%%Index Coding, Side information.
%%\end{IEEEkeywords}

\IEEEpeerreviewmaketitle
\section{Introduction}
\label{intro}
%The explosion of mobile applications and the expanded reach of mobile connectivity to an increasing number of end-users has led to network traffic congestion, particularly during peak traffic hours. In contrast, the resources are often underutilized during off-peak hours. Exploiting low-cost and abundant local memory to store contents at the user end proactively 

Caching is an effective way to utilize channel resources during off-peak periods and ease the burden of high network load at times of heavy demand. The idea of coded caching was introduced in the work of Maddah-Ali and Niesen \cite{MaN}, in which it is shown that by the joint design of storage and delivery policies, it is possible to achieve significant gain over conventional uncoded caching. The content delivery network in \cite{MaN} consists of a central server with a library of $N$ files, and a set of $K$ users with dedicated caches. However, the  capacity $M$ of the caches is limited. During off-peak hours, the server stores some of the file contents in the caches such that in the peak hour when the users demand files, the caches can help the server meet the user demands. The goal of the coded caching problem is the joint design of the placement phase (filling the caches in the off-peak period) and the delivery phase (server transmission in the peak traffic period) with a reduced transmission rate in the delivery phase. For a given $K,N$ and $M$, a coded caching scheme with uncoded placement and coded delivery was proposed in \cite{MaN}. By deriving a lower bound on the optimal rate-memory trade-off,  it is also shown that the rate achieved by the proposed scheme is within a constant multiplicative factor from the information-theoretic optimal rate-memory trade-off. Later in \cite{STC1,WLG,WBW,GhR}, better lower bounds were derived on the optimal rate-memory trade-off which significantly tightened the optimality gap.

However, in practical scenarios such as in cellular networks, users can have access to multiple caches when their coverage areas overlap. Motivated by the heterogeneous cellular architecture, a generalization of the set-up in \cite{MaN} is studied in \cite{HKD}, where each user has access to $L$ consecutive caches in a cyclic wrap-around manner (See Figure \ref{MACC}), instead of accessing a dedicated (single) cache as in \cite{MaN}. In \cite{HKD}, a lower bound on the optimal rate-memory trade-off for the multi-access coded caching (MACC) scheme is derived and provided achievability results. Later, the authors of \cite{ReK,SPE,ReK2,CLWZC,SaR,SaR2,MaR} came up with different MACC schemes, all restricted to uncoded placement.
\subsection{Contributions}
In this paper, we study the same multi-access set-up in \cite{ReK,SPE,ReK2,CLWZC,SaR,SaR2,MaR}, and make the following technical contributions.
\begin{itemize}
	\item A lower bound on the optimal rate-memory trade-off of MACC scheme is derived using cut-set arguments (Section \ref{mainresults}: Theorem \ref{cutset}). 
	\item An improved lower bound on the optimal rate-memory trade-off of MACC scheme is derived using non-cut-set arguments (Section \ref{mainresults}: Theorem \ref{lowerbound}).
	\item The proposed lower bound (in Theorem \ref{lowerbound}) is tighter than the previously known bounds in \cite{HKD} and \cite{HKD2} (Section \ref{mainresults}: Lemma \ref{better}).
	%\item Coded placement technique is introduced in the multi-access coded caching set-up for the first time. And we show that using coded placement we can significantly improve the rate, especially in the lower memory regime when $N\leq K$ (Section \ref{mainresults}: Theorem \ref{codedlbyk}).  
	%\item For some specific memory regimes, the exact optimality is shown by giving achievable schemes that match with the proposed lower bound (Section \ref{mainresults}: Lemma \ref{generalcfl}, Lemma \ref{generalSPE}, Lemma \ref{L=K-1}). 
\end{itemize}

\subsection{Notations}
For a positive integer $m$, $[m]$ denotes the set $ \left\{1,2,\hdots,m\right\}$. For two positive integers $m,n$ such that $m\leq n$, $[m:n] = \{m,m+1,\hdots,n\}$. For integers $m,n\leq K$, 
\begin{equation*}
	[m:n]_K =
	\begin{cases}
		\left\{m,m+1,\hdots,n\right\} & \text{if } m\leq n.\\
		\left\{m,m+1,\hdots,K,1,\hdots,n\right\} & \text{if } m>n.
	\end{cases}   
\end{equation*}
For any two integers, $i$ and $K$, 
\begin{equation*}
<i>_K =
\begin{cases}
i\text{ }(mod\text{ }K) & \text{if $i$ $(mod$ $K) \neq0$. }\\
K & \text{if $i$ $(mod$ $K) =0$.}
\end{cases}   
\end{equation*}
%For any two integers, $i$ and $K$, 
%\begin{equation*}
%	<i>_K =
%	\begin{cases}
%		i\text{ }(mod\text{ }K) & \text{if $i$ $(mod$ $K) \neq0$. }\\
%		K & \text{if $i$ $(mod$ $K) =0$.}
%	\end{cases}   
%\end{equation*}
For two positive integers $a$ and $b$, $\text{gcd}(a,b)$ represents the greatest common divisor of $a$ and $b$. Also, for $x\in \mathbb{R}$, $(x)^+=\max(0,x)$. The set of random variables $\{Y_a,Y_{a+1},\dots,Y_b\}$ is denoted as $Y_{[a:b]}$.

\section{System Model}
\label{systemmodel}
\begin{figure}[t]
	\begin{center}
		\captionsetup{justification = centering}
		\includegraphics[width = 0.8\columnwidth]{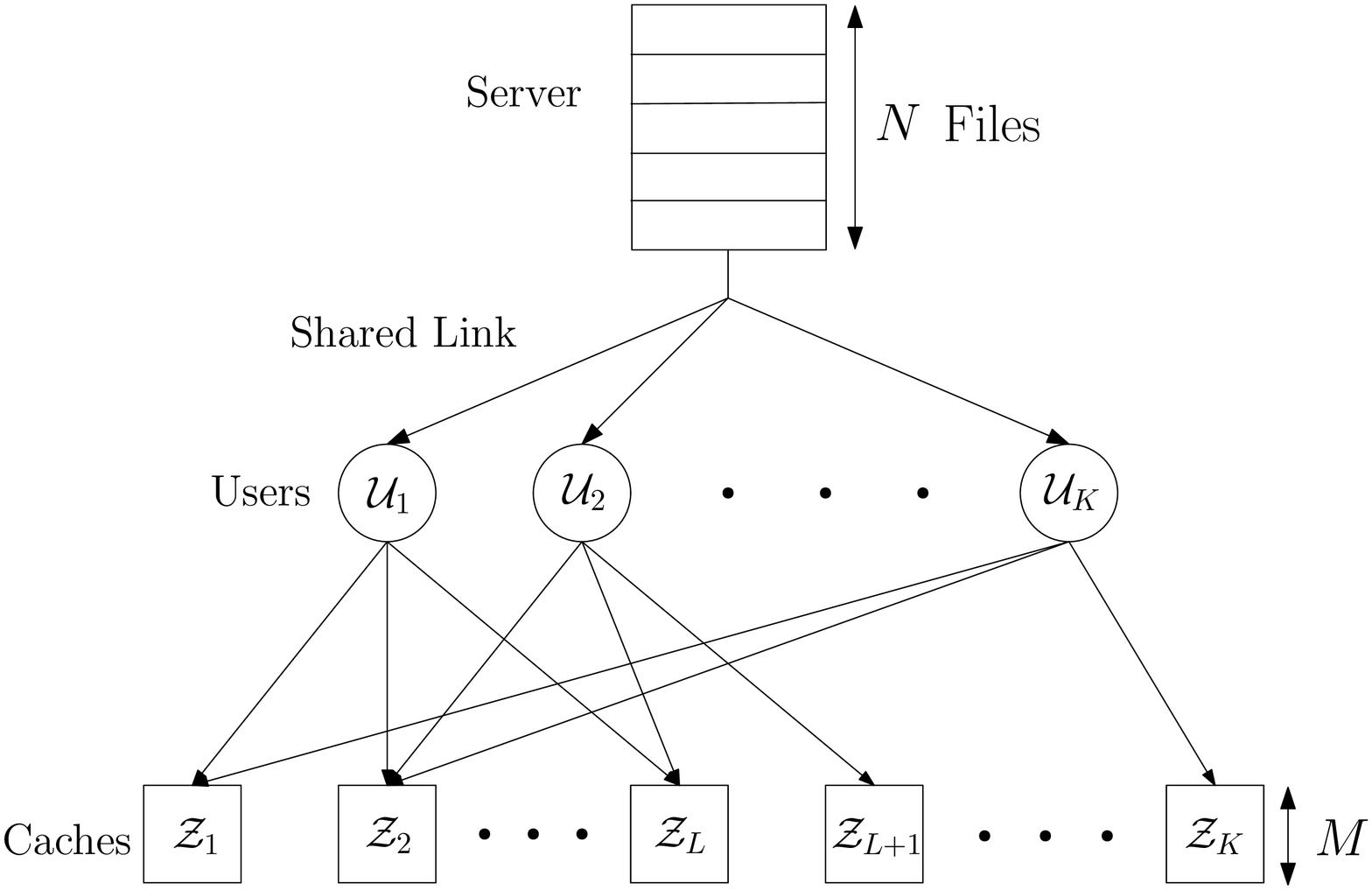}
		\caption{$(K,L,N)$ MACC Network. }
		\label{MACC}
	\end{center}
\end{figure}

We consider the multi-access network illustrated in Fig. \ref{MACC}. The system model consists of a central server and $K$ users.
The central server is having a library of $N$ independent files, $W_{[1:N]}$ each of size 1 unit ($F$ bits). The server is connected to a set of $K$ users $\mathcal{U}_{[1:K]}\triangleq \{\mathcal{U}_1,\mathcal{U}_2,\dots,\mathcal{U}_K\}$ through an error-free shared link. There are $K$ caches $\mathcal{Z}_{[1:K]}\triangleq \{\mathcal{Z}_1,\mathcal{Z}_2,\dots,\mathcal{Z}_K\}$ in the system, each of size $M$ units where $0\leq M \leq N$. Each user can access $L$ caches in a cyclic wrap-around manner. The set of caches that can be accessed by user $\mathcal{U}_k$ is denoted as $\mathcal{Z}_{\mathcal{L}_k}$, where $\mathcal{L}_k =[k:<k+L-1>_K]_K$. A system under this setting is called the $(K,L,N)$ MACC network. The coded caching problem under this model has been discussed in \cite{HKD,SPE,ReK,SaR,MaR,ReK2,CLWZC,SaR2}.

The $(K,L,N)$ MACC scheme works in two phases: the placement phase and the delivery phase. In the placement phase, the server stores some of the file contents in the caches without knowing the user demands. The placement can be either coded or uncoded. By uncoded placement, we mean that files will be split into subfiles and kept in the caches as such, while coded placement means that coded combinations of the subfiles are allowed to be kept in the caches. The content stored in cache $\mathcal{Z}_k$ is denoted as $Z_k$. The cache contents are the functions of the files. 
%That is, for every $k\in [K]$,
%\begin{equation}
%\label{cachefnoffiles}
%H(Z_k|W_{[1:N]}) = 0.
%\end{equation}
In the delivery phase, each user requests a file from the server. Let $D_k$ be the random variable denoting the demand of $\mathcal{U}_k$. The random variables $D_{[1:K]}$ are all uniformly and independently distributed over the set $[N]$. Let $\mathbf{d} = (d_1,d_2,\hdots,d_K)$ be the demand vector, where $d_k$ is a realization of $D_k$, for all $k\in [K]$. Corresponding to the demand vector $\mathbf{d}$, the server makes a broadcast transmission $X$ of size $R$ units. The non-negative real number $R$ is said to be the rate of transmission. The broadcast transmission $X$ is a function of the files $W_{[1:N]}$.% That is,
%\begin{equation}
%\label{txnfnoffiles}
%H(X|W_{[1:N]}) = 0.
%\end{equation}
Using the transmission $X$ and the contents in the set of caches $\mathcal{Z}_{\mathcal{L}_k}$ ($\triangleq \{\mathcal{Z}_k,\mathcal{Z}_{<k+1>_K},\dots,\mathcal{Z}_{<k+L-1>_K}$\}), user $\mathcal{U}_k$ should be able to decode the demanded file $W_{d_k}$. That is, for every $k\in [K]$, 
\begin{equation}
	\label{Correctness}
	\text{[Decodability Condition]  \hspace{0.15cm}   } H(W_{d_k}|Z_{\mathcal{L}_k},D_k=d_k,X)=0,\text{  } 
\end{equation}
where $Z_{\mathcal{L}_k}$ denotes the contents stored in the caches $\mathcal{Z}_{\mathcal{L}_k}$.
\begin{defn}	
For the $(K,L,N)$ MACC scheme, a memory-rate pair $(M,R)$ is said to be achievable if the scheme for cache memory $M$ satisfies the decodability condition \eqref{Correctness} with a rate less than or equal to $R$ for every possible demand vector (all possible realizations of $D_{[1:K]}$).
	
The optimal rate-memory trade-off is defined as, 
\begin{equation}
		\label{OptimalTradeoff}
		R^*(M) = \inf \left\{R: (M,R) \text{ is achievable}\right\}.
\end{equation}
\end{defn}

In this paper, we derive an improved lower bound on $R^*(M)$, which is tighter than the lower bounds in \cite{HKD} and \cite{HKD2}.
\section{Preliminaries}
\label{preliminaries}
In this section, we introduce the useful definitions and results that we will make use of in the subsequent sections. 

The following lemma will be useful in deriving our lower bound on $R^*(M)$.
\begin{lem}[Sliding-Window Subset Entropy Inequality, Theorem 3 in \cite{JML}]
	\label{Entropy}
	Given $K$ random variables $(Y_1,Y_2,\hdots,Y_K)$, we have, for every $s\in \{1,2,\hdots,K-1\}$,
\small{	\begin{equation}
		\frac{1}{s}\sum_{i=1}^K H(Y_i,\hdots,Y_{<i+s-1>_K})\geq \frac{1}{s+1}\sum_{i=1}^K H(Y_i,\hdots,Y_{<i+s>_K}).
	\end{equation}}
\end{lem}

Next, we see two existing lower bound results for MACC scheme, in the literature.
\begin{lem}[Lemma 5 in \cite{HKD}]
	\label{lowerHKD}
Consider the $(K,L,N)$ MACC scheme. Let $b\in \mathbb{N}^+$ and $t\in [K]$. Furthermore, let $s\in \mathbb{N}^+$ such that $st \in [L:\floor{\frac{K}{2}}]$. Then,
\begin{equation}
	R^*(M)\geq \lambda\cdot \min\left\{st-L+1, \frac{N}{sb}\right\}-\frac{t}{b}M
\end{equation}
where $\lambda = 1$ if $st=L$ and $\lambda =\frac{1}{2}$ if $st>L$.
\end{lem}
\begin{lem}[Lemma 3 in \cite{HKD2}]
	\label{cutsetHKD}
	For the $(K,L,N)$ MACC scheme,
	\begin{equation}
		R^*(M) \geq \max_{s\in \{1,2,\hdots,\min(K,N)\}} \left(s- \frac{s+L-1}{\floor{N/s}}M\right).
	\end{equation}
\end{lem}

Lemma \ref{cutsetHKD} gives a tighter lower bound on $R^*(M)$ compared to the lower bound in Lemma \ref{lowerHKD}. 

\section{Main Results and Discussions}
\label{mainresults}
In this section, first, we present a lower bound on $R^*(M)$  of the $(K,L,N)$ MACC scheme based on cut-set arguments. Then, we give an improved lower bound on $R^*(M)$ using non cut-set arguments.

\begin{thm}[Cut-set lower bound on $R^*(M)$]
	\label{cutset}
	For the $(K,L,N)$ MACC scheme, 
	\begin{equation}
		\small
		R^*(M) \geq \max_{s\in \{1,2,\hdots,\min(K,N)\}} \left(s- \frac{\min(s+L-1,K)}{\floor{N/s}}M\right).
	\end{equation}
\end{thm}
Proof of Theorem \ref{cutset} is given in Section \ref{App1}.\hfill $\blacksquare$

A similar, cut-set based approach is adopted to derive the lower bound in Lemma \ref{cutsetHKD} (\cite{HKD2}). However, the lower bound in Theorem \ref{cutset} is tighter than the lower bound in Lemma \ref{cutsetHKD} (This can be verified by considering values of $s$ such that $\min(s+L-1,K)=K$). We further improve this lower bound in Theorem \ref{lowerbound}. As an example, we look into the $(K=3,L=2,N=3)$ coded caching set-up. From Theorem \ref{cutset}, we get $R^*(M)\geq 3(1-M)$ and $R^*(M)\geq 1-\frac{2M}{3}$. Along with these two, we find one more lower bound such that all the three together characterize the improved lower bound on $R^*(M)$ of the $(K=3,L=2,N=3)$ MACC scheme. This improved lower bound is, in fact, the optimal rate-memory trade-off of the $(K=3,L=2,N=3)$ multi-access coded caching schemes (see Appendix \ref{App3}).
\begin{exmp}
	\label{example2}
	Consider the $(K=3,L=2,N=3)$ MACC scheme. Consider three demand vectors $\mathbf{d}_1 = (1,2,3)$, $\mathbf{d}_2 = (2,3,1)$ and $\mathbf{d}_3 = (3,1,2)$. Corresponding to $\mathbf{d}_1$, the server makes the transmission $X_1$. Similarly, corresponding to $\mathbf{d}_2 $ and $\mathbf{d}_3$, the server makes the transmissions $X_2$ and $X_3$, respectively. Using the transmissions $X_1$, $X_2$ and $X_3$, and the cache contents $Z_1$ and $Z_2$, all the files can be decoded. Therefore, we have,
	\begin{subequations}
		\begin{align}
			3&\leq H(Z_{[1:2]},X_{[1:3]})\notag\\
			&=    H(Z_{[1:2]})+H(X_{[1:3]}|Z_{[1:2]})\notag\\
			&\leq 2M+H(X_{1}|Z_{[1:2]})+H(X_{[2:3]}|Z_{[1:2]},X_{1})\notag\\
			&\leq 2M+H(X_{1})+H(X_{[2:3]}|Z_{[1:2]},X_{1},W_1)\label{ex2eqd}\\
			&\leq 2M+R^*(M) +H(X_{[2:3]},Z_3|Z_{[1:2]},X_{1},W_1)\notag\\
			&\leq 2M+R^*(M)+H(Z_3|Z_{[1:2]},X_{1},W_1)\notag\\
			&\qquad+H(X_{[2:3]}|Z_{[1:3]},X_{1},W_1)\notag\\
			&\leq 2M+R^*(M)+H(Z_3|Z_{[1:2]},X_{1},W_1)\notag\\
			&\qquad+H(X_{[2:3]}|Z_{[1:3]},X_{1},W_{[1:3]})\label{ex2eqg}\\
			&\leq 2M+R^*(M)+H(Z_3|Z_{[1:2]},X_{1},W_1)\label{ex2eqh}.
		\end{align}
	\end{subequations}
	Using $X_{1}$ and $Z_{[1:2]}$, file $W_1$ can be decoded, therefore \eqref{ex2eqd} follows. Similarly, using  $X_{1}$ and $Z_{[1:3]}$, files $W_1,W_2$ and $W_3$ can be decoded, hence \eqref{ex2eqg} follows. Equation \eqref{ex2eqh} follows from the fact that given the files $W_{[1:3]}$, there is no more uncertainty in the transmissions $X_{2}$ and $X_{3}$. Now, 
	\begin{align*}
		H(Z_3|Z_{[1:2]},X_{1},W_1) &\leq H(Z_3|Z_{[1:2]},W_1)\\
		&= H(Z_{[1:3]}|W_1)-H(Z_{[1:2]}|W_1).
	\end{align*}
	Therefore, we have,
	\begin{equation}
		3\leq 2M+R^*(M)+H(Z_{[1:3]}|W_1)-H(Z_{[1:2]}|W_2).
		\label{ex2eq1}
	\end{equation}
	Instead of $Z_{[1:2]}$ and $X_{[1:3]}$, by considering $Z_{[2:3]}$ and $X_{[1:3]}$, we will get a similar inequality as follows: 
	\begin{equation}
		3\leq 2M+R^*(M)+H(Z_{[1:3]}|W_1)-H(Z_{[2:3]}|W_1).
		\label{ex2eq2}
	\end{equation}
	Similarly, considering $Z_{[3:1]_3}$ and $X_{[1:3]}$ will give,
	\begin{equation}
		3\leq 2M+R^*(M)+H(Z_{[1:3]}|W_1)-H(Z_{[3:1]_3}|W_1).
		\label{ex2eq3}
	\end{equation}
	Adding \eqref{ex2eq1},\eqref{ex2eq2} and \eqref{ex2eq3}, and normalizing with 3,
	\begin{equation}
		\small
		3\leq 2M+R^*(M)+H(Z_{[1:3]}|W_1)-\frac{1}{3}\sum_{i=1}^3H(Z_{[i:<i+1>_3]_3}|W_1).
		\label{ex2eq4}
	\end{equation}
	By using Lemma \ref{Entropy}, we get,
	\begin{equation}
		\sum_{i=1}^3 H(Z_{[i:<i+1>_3]_3}|W_1)\geq 2H(Z_{[1:3]}|W_1)
		\label{ex2eq5}
	\end{equation}
	Substituting \eqref{ex2eq5} in \eqref{ex2eq4},
	\begin{subequations}
		\begin{align}
			3&\leq 2M+R^*(M)+\frac{1}{3}H(Z_{[1:3]}|W_1)\notag\\
			&\leq 2M+R^*(M)+\frac{1}{3}H(Z_{[1:3]},W_{[2:3]}|W_1)\notag\\
			&\leq 2M+R^*(M)+\frac{1}{3}\left(H(W_{[2:3]}|W_1)+H(Z_{[1:3]}|W_{[1:3]})\right) \label{ex2eq3c}\\
			&\leq 2M+R^*(M)+\frac{1}{3}H(W_{[2:3]})\notag\\
			&\leq 2M+R^*(M)+\frac{2}{3}\label{ex2eq6}	
		\end{align}
	\end{subequations}
	where \eqref{ex2eq3c} results from the fact that, given the files $W_{[1:3]}$, there is no uncertainty in $Z_{[1:3]}$.
	
	Rearranging the terms will give the lower bound on $R^*(M)$ as follows,
	\begin{equation}
		R^*(M)\geq \frac{7}{3}-2M.
		\label{ex2eq7}
	\end{equation}
		\begin{figure}[t]
		\begin{center}
			\captionsetup{justification = centering}
			\captionsetup{font=small,labelfont=small}
			\includegraphics[width = 0.8\columnwidth]{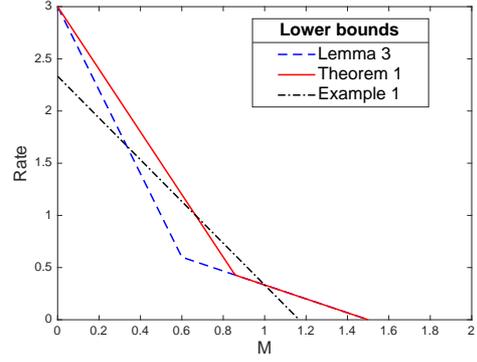}
			\caption{$(K=3,L=2,N=3)$ MACC scheme: lower bounds on $R^*(M)$}
			\label{exam2}
		\end{center}
	\end{figure}
\end{exmp}
For the $(K=3,L=2,N=3)$ MACC scheme, the lower bound on $R^*(M)$ given by  Lemma \ref{cutsetHKD}, Theorem \ref{cutset} along with \eqref{ex2eq7} is shown in Figure \ref{exam2}. Note that, for $\frac{2}{3}\leq M\leq 1$, Example \ref{example2} improves the lower bound. The following theorem generalizes Example \ref{example2}, and gives an improved lower bound on $R^*(M)$.

%\subsection{Lower Bound}
\begin{thm}
	\label{lowerbound}
	For the $(K,L,N)$ MACC scheme,
\small{	\begin{align}
	\label{lowereqn}
R^*&(M)
\geq \max_{\substack{s\in \{1,2,\hdots,K\} \\ l\in \left\{1,2,\hdots,\ceil{N/s}\right\}}} \notag\\ &\left(\frac{1}{l}\left\{N-\left(1-\frac{p}{K}\right)(N-ls)^+ - (N-lK)^+-pM\right\}\right) 
	\end{align}}  
	where $p = \min(s+L-1,K)$.
\end{thm}

Proof of Theorem \ref{lowerbound} is given in Section \ref{proofoflowerbound}. \hfill $\blacksquare$

The lower bound in Theorem \ref{lowerbound} has two parameters: $s$, which is associated to the number of users, and the parameter $l$ connected to the number of transmissions. Compared to the lower bound in Theorem \ref{cutset}, the parameter $l$ gives more flexibility to Theorem \ref{lowerbound}. To derive the lower bound, we consider $s$ users, where $s\leq K$. These $s$ users together access $p$ caches. From $\ceil{N/s}$ transmissions, all the $N$ files can be decoded at the user-end (considering only those $s$ users). We separate those $\ceil{N/s}$ transmissions into the first $l$ transmissions and the remaining $\ceil{N/s}-l$ transmissions and bound the uncertainty in the latter with the help of Lemma \ref{Entropy}. This approach is similar to the approach in \cite{STC1} in which a lower bound on the optimal rate-memory trade-off of the dedicated coded caching scheme (setting introduced in \cite{MaN}) is derived. 

\begin{figure}[h!]
	\begin{subfigure}[h]{0.5\textwidth}
		\centering
		\captionsetup{justification = centering}
		\captionsetup{font=small,labelfont=small}
		\includegraphics[width = 0.7\textwidth]{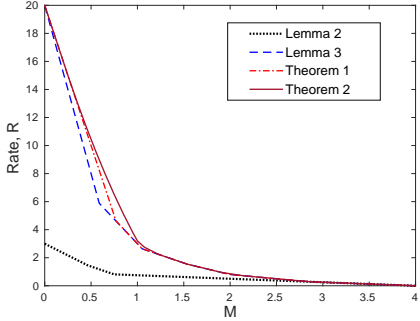}
		\caption{$(K=20,L=5,N=20)$ MACC scheme}
		\label{result1}
	\end{subfigure}
	\hfill
	\begin{subfigure}[h]{0.5\textwidth}
		\centering
		\captionsetup{justification = centering}
		\captionsetup{font=small,labelfont=small}
		\includegraphics[width = 0.7\textwidth]{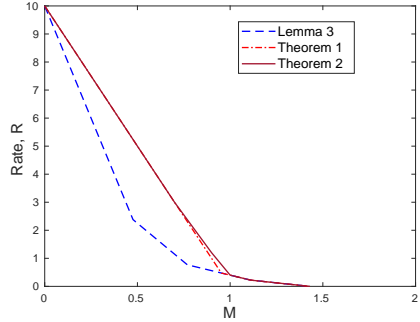}
		\caption{$(K=10,L=7,N=10)$ MACC scheme}
		\label{result2}
	\end{subfigure}
	\caption{Lower bounds on $R^*(M)$}
	\label{results1}
\end{figure}

For different values of $K,L$ and $N$, the lower bounds on $R^*(M)$ in Lemma \ref{lowerHKD}, Lemma \ref{cutsetHKD}, Theorem \ref{cutset} and Theorem \ref{lowerbound} are given in Figure \ref{results1}. The lower bound in Lemma \ref{lowerHKD} is applicable only when $L\leq \frac{K}{2}$. Also, we have already shown that the lower bound in Theorem \ref{cutset} is tighter than the lower bound in Lemma \ref{cutsetHKD}. In the proof of Lemma \ref{better}, we analytically show that for any values of $K,L$ and $N$, the lower bound in Theorem \ref{lowerbound} is at least as good as the lower bound in Theorem \ref{cutset}. Also, from Figure \ref{results1}, it is clear that the lower bound in Theorem \ref{lowerbound} is, in fact, tighter than the lower bound in Theorem \ref{cutset}.

\begin{lem}
	\label{better}
	Let 
	\begin{equation}
	\label{bettereqn1}
	R_{cut-set}(M) = \max_{s\in \{1,2,\hdots,\min(K,N)\}} \left(s- \frac{p}{\floor{N/s}}M\right)
	\end{equation}
and 
\small{\begin{align}
\label{bettereqn2}
R_{new}&(M)
\geq \max_{\substack{s\in \{1,2,\hdots,K\} \\ l\in \left\{1,2,\hdots,\ceil{N/s}\right\}}} \notag\\ &\left(\frac{1}{l}\left\{N-\left(1-\frac{p}{K}\right)(N-ls)^+ - (N-lK)^+-pM\right\}\right) 
\end{align}}
\noindent where $p = \min(s+L-1,K)$. Then, for all $M\in [0,\frac{N}{L}]$, we have, $R_{new}(M)\geq R_{cut-set}(M)$.
\end{lem}

Proof of Lemma \ref{better} is given in Appendix \ref{App2}. \hfill $\blacksquare$

In Figure \ref{fig}, the rate-memory trade-off of different multi-access schemes is given along with the proposed lower bound on $R^*(M)$. By substituting $s=1$ and $l =N$ in \eqref{lowereqn}, we get, $R^*(M)\geq 1-\frac{LM}{N}$. It says that, $R^*(\frac{N}{L})\geq 0$. Restricted to uncoded placement, the minimum value of $M$ required for achieving rate $R = 0$ is $\ceil{\frac{K}{L}}\frac{N}{K}$, where $\ceil{\frac{K}{L}}\frac{N}{K}>\frac{N}{L}$ when $L$ does not divide $K$. This means, in general, the optimal rate restricted to uncoded placement at $M=\frac{N}{L}$ is a non-zero value. At the same time, using coded placement, rate $R=0$ is achievable at $M=\frac{N}{L}$ itself, which is a clear indication that by making use of coded placement, it is possible to reduce the rate further compared to the rate achieved by the schemes in \cite{HKD,ReK,SPE,ReK2,CLWZC,SaR,SaR2,MaR}.  

\begin{figure*}\centering
	\subfloat[$K=10,L=6,N=10$]{\label{a}\includegraphics[width=.33\linewidth]{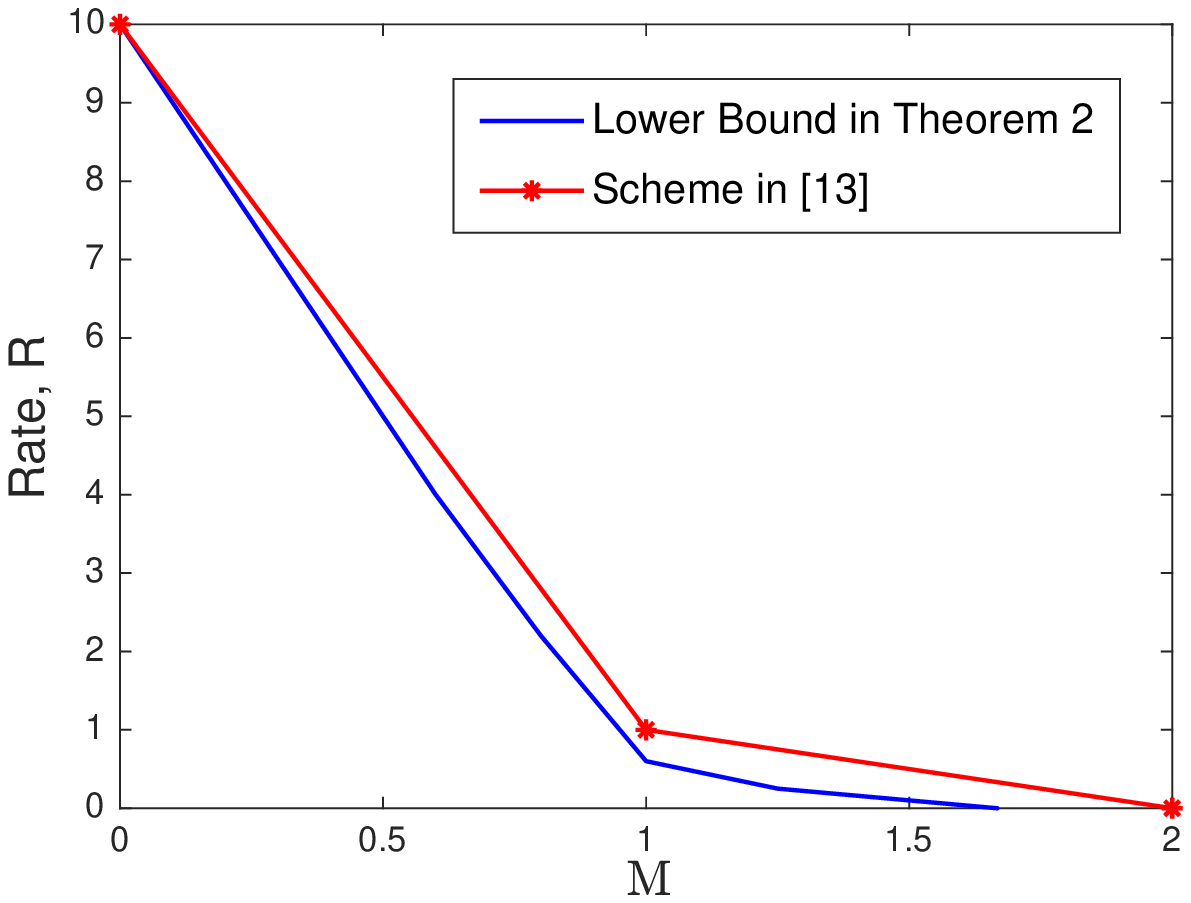}}\hfill
	\subfloat[$K=11,L=3,N=11$ ]{\label{b}\includegraphics[width=.333\linewidth]{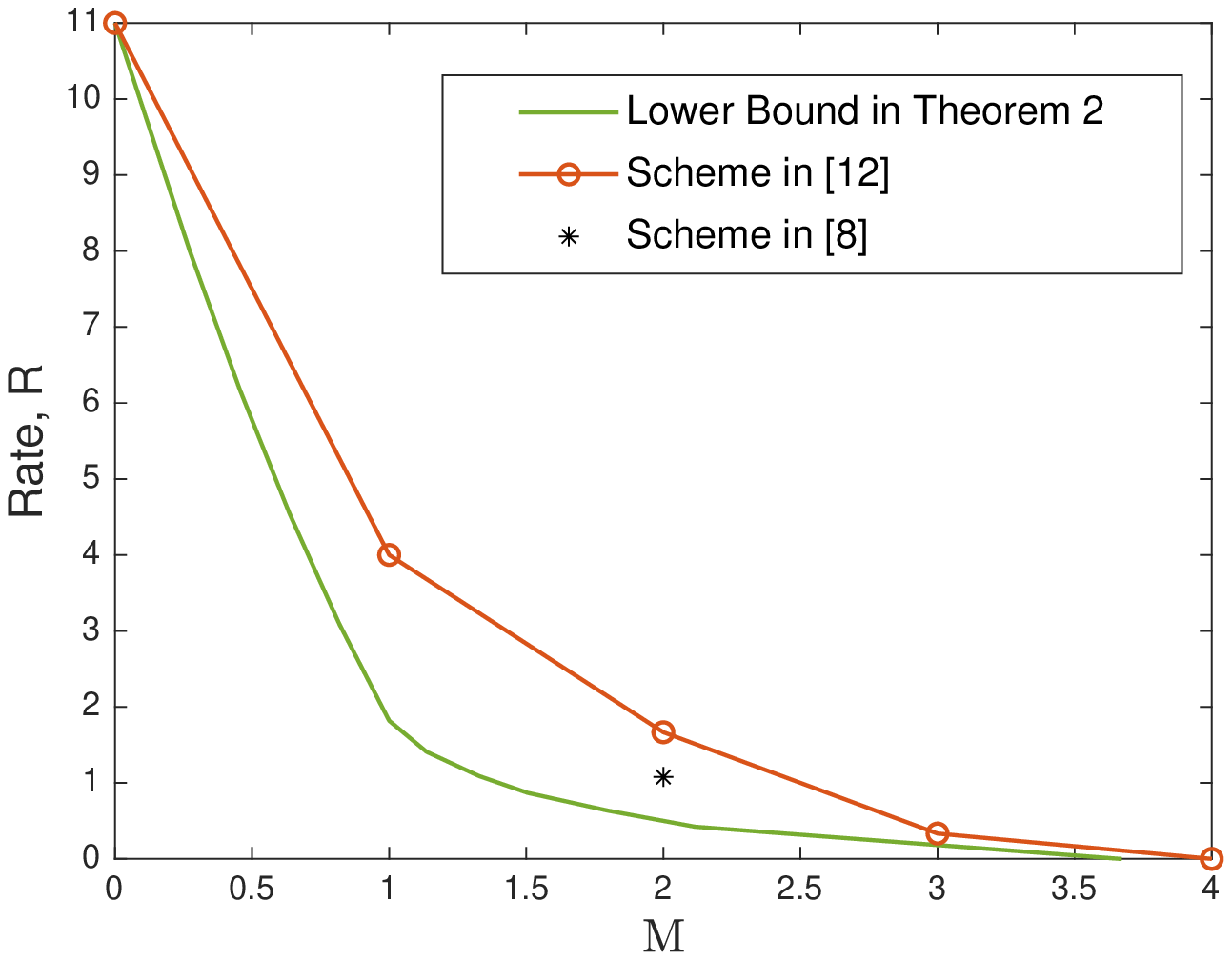}}\hfill
	\subfloat[$K=10,L=3,N=10$]{\label{c}\includegraphics[width=.33\linewidth]{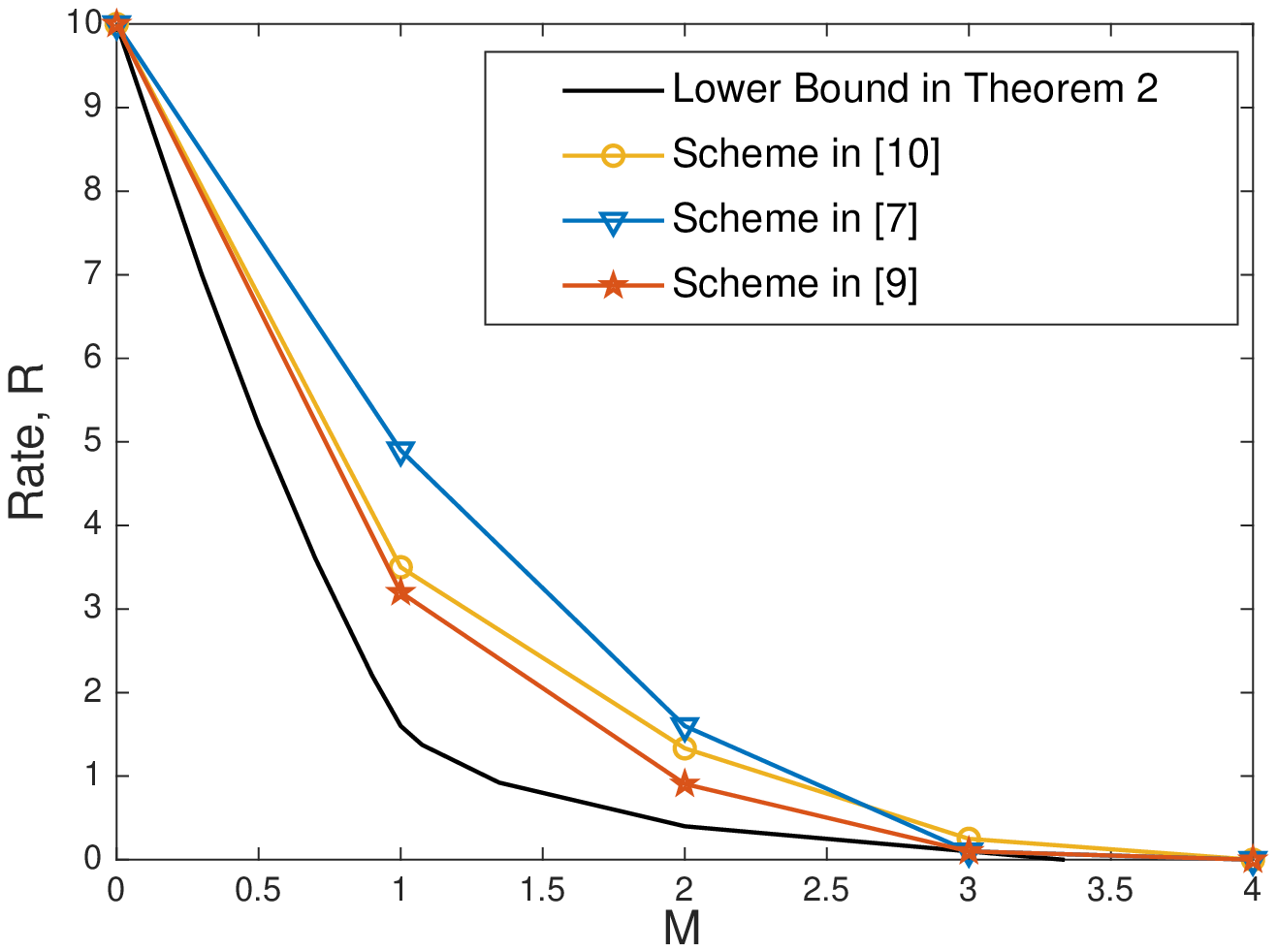}}
	\caption{MACC schemes in \cite{ReK,ReK2,SPE,CLWZC,SaR2,MaR}, and the proposed lower bound}
	\label{fig}
\end{figure*}

\section{Proof of Theorem \ref{cutset} and Theorem \ref{lowerbound}}
\label{proofs}
In this section, we derive the lower bounds on $R^*(M)$ in Theorem \ref{cutset} and Theorem \ref{lowerbound}.
\subsection{Proof of Theorem \ref{cutset}}
\label{App1}	
In this section, we derive a cut-set based lower bound on $R^*(M)$. Let $s$ be an integer such that $s\in \{1,2,\hdots,\min(K,N)\}$. Let $p = \min(s+L-1,K)$. Consider the caches $\mathcal{Z}_{[1:p]}$. Corresponding to the demand vector $(1,2,\hdots,s,d_{s+1},\hdots,d_K)$ (where the first $s$ users request for the specified $s$ files, and the remaining $K-s$ users' requests can be arbitrary files), the server makes transmission $X_1$. Using $X_{1}$ and the cache contents $Z_{[1:p]}$, the files $W_{[1:s]}$ can be decoded. Similarly, using the same cache contents and the transmission corresponding the demand vector $(s+1,s+2,\hdots,2s,d_{s+1},\hdots,d_K)$ (where the first $s$ users request for the specified $s$ files, and the remaining $K-s$ users' requests can be arbitrary files), the files $W_{[s+1:2s]}$ can be decoded. Therefore, considering $\floor{N/s}$ such demand vectors, there will be $\floor{N/s}$ transmissions, $X_{[1:\floor{N/s}]}$.
Using the contents of the caches $\mathcal{Z}_{[1:p]}$, and from the transmissions $X_{[1:\floor{N/s}]}$, the files $W_{[1:s\floor{N/s}]}$ can be decoded. Therefore, 
\begin{subequations}
	\begin{align*}
		s\floor{N/s} &\leq H(X_{[1:\floor{N/s}]},Z_{[1:p]})\leq \floor{N/s} R^*(M)+ pM.
	\end{align*}
\end{subequations}
Upon rearranging, we get,
\begin{subequations}
	\begin{align*}
		R^*(M) &\geq \frac{s\floor{N/s}-pM}{\floor{N/s}}= s- \frac{p}{\floor{N/s}}M.
	\end{align*}
\end{subequations}
Optimizing over all possible choices of $s$, we obtain the lower bound on $R^*(M)$ as,
\begin{equation*}
	R^*(M) \geq \max_{s\in \{1,2,\hdots,\min(N,K)\}} \left(s- \frac{\min(s+L-1,K)}{\floor{N/s}}M\right).
\end{equation*}

This completes the proof of Theorem \ref{cutset}.\hfill $\blacksquare$

\subsection{Proof of Theorem \ref{lowerbound}}
\label{proofoflowerbound}
The server has $N$ files $W_{[1:N]}$. Consider the users $\mathcal{U}_{[1:s]}$ for some $s\in [K]$. These $s$ users together access the caches $\mathcal{Z}_{[1:p]}$ where $p \triangleq \min(s+L-1,K)$. That is, $\mathcal{L}_1 \cup \mathcal{L}_2 \cup \dots \cup \mathcal{L}_s=[p]$. Consider transmission $X_1$ corresponding to the demand vector $\mathbf{d}_1 = (1,2,\dots,s,d_{s+1},\dots,d_K)$, where the first $s$ requests are for distinct files, and the last $K-s$ requests can be for arbitrary files. Similarly, the transmission $X_2$ is corresponding to the demand vector $\mathbf{d}_2 = (s+1,s+2,\dots,2s,d_{s+1},\dots,d_K)$. Finally, the transmission $X_q$ is corresponding to the demand vector $\mathbf{d}_q = ((q-1)s+1,(q-1)s+2,\dots,N,d_{N-(q-1)s+1},\dots,d_K)$, where $q \triangleq \ceil{N/s}$. Note that, all the transmissions are functions of the files $W_{[1:N]}$ alone. Using the cache contents $Z_{[1:p]}$ and the transmissions $X_{[1:q]}$, the files $W_{[1:N]}$ can be decoded. Therefore, we have,
\begin{subequations}
	\begin{align}
	N& = H(W_{[1:N]})\leq H(Z_{[1:p]},X_{[1:q]}) \notag\\
	&\leq    H(Z_{[1:p]})+H(X_{[1:q]}|Z_{[1:p]})\notag\\
	&\leq pM+H(X_{[1:l]}|Z_{[1:p]})+H(X_{[l+1:q]}|Z_{[1:p]},X_{[1:l]})\notag\\
	&\leq pM+H(X_{[1:l]})+H(X_{[l+1:q]}|Z_{[1:p]},X_{[1:l]},W_{[1:ls]})\label{Wls}\\
	&\leq pM+lR^*(M)\notag\\
	&\qquad+H(X_{[l+1:q]},Z_{[p+1:K]}|Z_{[1:p]},X_{[1:l]},W_{[1:ls]})\label{lRstar}\\
	&\leq pM+lR^*(M)+\underbrace{H(Z_{[p+1:K]}|Z_{[1:p]},X_{[1:l]},W_{[1:ls]})}_{\triangleq \mu}\notag\\
	&\qquad +\underbrace{H(X_{[l+1:q]}|Z_{[1:K]},X_{[1:l]},W_{[1:ls]})}_{\triangleq \psi} \label{mupsi}
	\end{align}
\end{subequations}
where \eqref{Wls} follows from the fact that the cache contents $Z_{[1:s]}$ with transmissions $X_{[1:l]}$ can decode the files $W_{[1:ls]}$. Equation \eqref{lRstar} results from bounding the entropy of $l$ transmissions by $lR^*(M)$, where each transmission is of rate $R^*(M)$.

Now, we will find an upper bound on $\mu$,
\begin{subequations}
	\begin{align}
	\mu &= H(Z_{[p+1:K]}|Z_{[1:p]},X_{[1:l]},W_{[1:ls]})\notag\\
	&\leq H(Z_{[p+1:K]}|Z_{[1:p]},W_{[1:ls]})\notag\\
	&\leq H(Z_{[1:K]}|W_{[1:ls]})-H(Z_{[1:p]}|W_{[1:ls]}).\label{mueqn}
	\end{align}
\end{subequations}
Considering the $p$-consecutive cache subsets of $\mathcal{Z}_{[1:K]}$ (all the sets of caches of the form $\mathcal{Z}_{[k:<k+p-1>_K]_K}$, for $k\in [K]$), we can obtain $K$ such inequalities as in \eqref{mueqn}. That is, for every $k\in [K]$, we have,
\begin{equation*}
\mu\leq  H(Z_{[1:K]}|W_{[1:ls]})-H(Z_{[k:<k+p-1>_K]_K}|W_{[1:ls]}).
\end{equation*}
Adding all the $K$ inequalities, and normalizing with $K$, we get,
\begin{equation}
\mu \leq H(Z_{[1:K]}|W_{[1:ls]})-\frac{1}{K}\sum_{k=1}^K H(Z_{[k:<k+p-1>_K]_K}|W_{[1:ls]})\label{mubound}.
\end{equation}
Next, consider the random variables $Z_k$ for all $k\in [K]$, and apply Lemma \ref{Entropy}. We get,
\begin{subequations}
	\begin{align}
	\frac{1}{p}&\sum_{k=1}^K H(Z_{[k:<k+p-1>_K]_K}|W_{[1:ls]}) \geq \notag\\
	&\qquad \frac{1}{K}\sum_{k=1}^K H(Z_{[k:<k+K-1>_K]_K}|W_{[1:ls]})\notag\\
	&= \frac{1}{K}\sum_{k=1}^K H(Z_{[1:K]}|W_{[1:ls]})= H(Z_{[1:K]}|W_{[1:ls]}).\label{mueqn2}
	\end{align}
\end{subequations}
Substituting \eqref{mueqn2} in \eqref{mubound}, we have,
\begin{subequations}
	\begin{align*}
	\mu &\leq H(Z_{[1:K]}|W_{[1:ls]})-\frac{1}{K}\sum_{k=1}^K H(Z_{[k:<k+p-1>_K]_K}|W_{[1:ls]})\\
	&\leq H(Z_{[1:K]}|W_{[1:ls]})-\frac{p}{K}H(Z_{[1:K]_K}|W_{[1:ls]})\\
	&=    \left(1-\frac{p}{K}\right)H(Z_{[1:K]}|W_{[1:ls]}).
	\end{align*}
\end{subequations}
Now, consider two cases a) if $ls\geq N$, then
\begin{align}
H(Z_{[1:K]}|W_{[1:ls]})=0\label{mueqn3}
\end{align}

and b) if $ls < N$, then, 
\begin{subequations}
	\begin{align}
	H(Z_{[1:K]}|W_{[1:ls]})&\leq H(Z_{[1:K]},W_{[ls+1:N]}|W_{[1:ls]})\\
	&\leq H(W_{[ls+1:N]}|W_{[1:ls]})\notag\\
	&\qquad +H(Z_{[1:K]}|W_{[1:N]})\\
	&\leq N-ls\label{mueqn4}
	\end{align}
\end{subequations}
where \eqref{mueqn3} and \eqref{mueqn4} follow from the fact that the cache contents are the functions of the files.

So, in general, we have,
\begin{align}
H(Z_{[1:K]}|W_{[1:ls]})\leq (N-ls)^+.
\end{align}
Therefore, we get an upper bound on $\mu$ as,
\begin{align}
\mu \leq \left(1-\frac{p}{K}\right)(N-ls)^+.\label{mu}
\end{align}
Now, we will find an upper bound on $\psi$, where,
\begin{align*}
\psi &= H(X_{[l+1:q]}|Z_{[1:K]},X_{[1:l]},W_{[1:ls]}).	
\end{align*}
We consider two cases a) if $N\leq lK$, then all the files $W_{[1:N]}$ can be decoded from $Z_{[1:K]}$ and $X_{[1:l]}$. Then the uncertainty in $X_{[l+1:q]}$ is zero. That is, when $N\leq lK$, 
\begin{align*}
\psi &= 0.	
\end{align*}
And b) if $N> lK$, we have,
\begin{subequations}
	\begin{align}
	\psi &= H(X_{[l+1:q]}|Z_{[1:K]},X_{[1:l]},W_{[1:ls]})\notag \\
	&= H(X_{[l+1:q]}|Z_{[1:K]},X_{[1:l]},W_{[1:lK]})\label{psieqn1}\\
	&\leq H(X_{[l+1:q]},W_{[lK+1:N]}|Z_{[1:K]},X_{[1:l]},W_{[1:lK]})\notag\\	
	&\leq H(W_{[lK+1:N]}|Z_{[1:K]},X_{[1:l]},W_{[1:lK]})\notag\\
	&\qquad+H(X_{[l+1:q]}|Z_{[1:K]},X_{[1:l]},W_{[1:N]})\notag\\
	&= H(W_{[lK+1:N]}|Z_{[1:K]},X_{[1:l]},W_{[1:lK]})\label{psieqn2}\\
	&\leq H(W_{[lK+1:N]})\leq N-lK.\notag
	\end{align}
\end{subequations}
Accessing the caches $\mathcal{Z}_{[1:K]}$, the transmissions $X_{[1:l]}$ will suffice to decode the files $W_{[1:lK]}$, \eqref{psieqn1} follows. Similarly, \eqref{psieqn2} follows from the fact that given $W_{[1:N]}$ there is no uncertainty in the transmissions. We have the upper bound on $\psi$,
\begin{align}
\psi \leq (N-lK)^+ \label{psi}.
\end{align}
Substituting \eqref{mu} and \eqref{psi} in \eqref{mupsi}, we get,
\begin{equation*}
N \leq pM+lR^*(M)+\left(1-\frac{p}{K}\right)(N-ls)^+ + (N-lK)^+.
\end{equation*} 
Upon rearranging the terms, and optimizing over all the possible values of $s$ and $l$, we have,
\begin{align*}
R^*&(M)
\geq \max_{\substack{s\in \{1,2,\hdots,K\} \\ l\in \left\{1,2,\hdots,\ceil{N/s}\right\}}} \notag\\ &\frac{1}{l}\left\{N-\left(1-\frac{p}{K}\right)(N-ls)^+ - (N-lK)^+-pM\right\} 
\end{align*}  
where $p = \min(s+L-1,K)$.

This completes the proof of Theorem \ref{lowerbound}.\hfill $\blacksquare$

\section{Conclusion}
In this work, we dealt with the MACC problem. We derived a lower bound on the optimal rate-memory trade-off of the MACC scheme using non-cut-set arguments. The proposed lower bound is tighter than the previously known lower bounds in the literature.

\begin{appendices}
	\label{appendix}

\section{Proof of Lemma \ref{better}}
\label{App2}
In \eqref{bettereqn2}, we substitute $l = \floor{N/s}$. Then,
\begin{align*}
R&_{new}(M)\geq \max_{s\in \{1,\hdots,K\}} \\
&\frac{N-\left(1-\frac{p}{K}\right)(N-\floor{N/s}s) - (N-\floor{N/s}K)^+-pM}{\floor{N/s}}.
\end{align*}
First, we consider the case with $\min(K,N)=N$. Then, we have,
\begin{subequations}
	\begin{align*}
	R_{new}(M)&\geq \max_{s\in \{1,\hdots,N\}} \left(s+\frac{\frac{p}{K}(N-s\floor{N/s})-pM}{\floor{N/s}}\right)\\
	&\geq \max_{s\in \{1,\hdots,N\}} \left(\frac{s\floor{N/s}-pM}{\floor{N/s}}\right)\geq R_{cut-set}(M).
	\end{align*}
\end{subequations}
Next, we consider the case with $\min(K,N)=K$. For each $s\in [1:K]$, we define
\begin{align*}
R&_{new}^s(M)\triangleq \\
&\frac{N-\left(1-\frac{p}{K}\right)(N-\floor{N/s}s)-(N-\floor{N/s}K)^+-pM}{\floor{N/s}}
\end{align*}
and
\begin{equation*}
R_{cut-set}^s(M)\triangleq \frac{s\floor{N/s}-pM}{\floor{N/s}}.
\end{equation*}
Now, we will show that, for every $s\in [1:K]$, 
\begin{equation}
\label{NgeqK}
R_{new}^s(M)\geq R_{cut-set}^s(M).
\end{equation}
Then we can say, when $N> K$, $R_{new}(M)\geq R_{cut-set}(M)$.

First, consider $s\in [K-L+1,K]$. Then, $p = \min(s+L-1,K) =K$. For every $s\in [K-L+1,K]$, we have,
\begin{subequations}
	\begin{align*}
	R_{new}^s(M)&\geq \frac{N-(N-\floor{N/s}K)^+-pM}{\floor{N/s}}\\
	&= \frac{\min(N,K\floor{N/s})-pM}{\floor{N/s}}\\
	&\geq \frac{s\floor{N/s}-pM}{\floor{N/s}}\geq R_{cut-set}^s(M).
	\end{align*}
\end{subequations}
For $s\in [1:K-L]$,
\begin{subequations}
	\begin{align*}
		R&_{new}^s(M)\geq \\
		&\frac{N-\left(1-\frac{p}{K}\right)(N-s\floor{N/s})-(N-\floor{N/s}K)^+-pM}{\floor{N/s}}\\
		&=	s+\frac{\frac{p}{K}(N-s\floor{N/s})-(N-\floor{N/s}K)^+-pM}{\floor{N/s}}.
	\end{align*}
\end{subequations}
If $(N-\floor{N/s}K)^+ = 0$, then 
\begin{subequations}
	\begin{align*}
	R_{new}^s(M)&\geq \frac{s\floor{N/s}+\frac{p}{K}(N-s\floor{N/s})-pM}{\floor{N/s}}\\
				&\geq \frac{s\floor{N/s}-pM}{\floor{N/s}}\geq R_{cut-set}^s(M).
	\end{align*}
\end{subequations}
That is, \eqref{NgeqK} holds for $s\in [1:K-L]$ such that $N-\floor{N/s}K\leq 0$. Note that, 
\begin{equation*}
	N-\floor{N/s}K\leq 0 \implies s\leq \floor[\Big]{\frac{N}{\ceil{N/K}}}.
\end{equation*}
For all $s\leq \floor[\Big]{\frac{N}{\ceil{N/K}}}$, $R_{new}^s(M)\geq R_{cut-set}^s(M)$.

Now, it remains to show that for $s\in \left[\floor[\Big]{\frac{N}{\ceil{N/K}}}+1:K-L\right]$, $R_{new}^s(M)\geq R_{cut-set}^s(M)$. Notice that, for those values of $s$, we have, $N/K>\floor{N/s}$. But, for all $s\leq K$, $N/K\leq N/s$. That is, for $s$ such that, $\floor{N/s}< N/K< N/s$, we have, $\floor{N/K} = \floor{N/s}$. Then,
	\begin{align*}
		R_{new}^s(M)\geq s+\frac{\frac{p}{K}(N-s\floor{N/s})+K\floor{N/s}-N-pM}{\floor{N/s}}.
	\end{align*}
For \eqref{NgeqK} to be true, $\frac{p}{K}(N-s\floor{N/s})+K\floor{N/s}-N \geq 0$. That implies,
\begin{equation}
	\label{smallset}
\frac{p}{K}(N-s\floor{N/s}) \geq N - K\floor{N/s}.	
\end{equation}
Consider the inequality,
\begin{equation}
	\label{largeset}
	\frac{s}{K}(N-s\floor{N/s}) \geq N - K\floor{N/s}.	
\end{equation} 
 If $s$ satisfies \eqref{largeset}, then $s$ will definitely satisfy \eqref{smallset}, since $p\geq s$. To satisfy \eqref{largeset}, 
 \begin{equation}
 	\label{Kands}
 	\frac{s}{K}\geq \frac{\frac{N}{\floor{N/K}}-K}{\frac{N}{\floor{N/K}}-s}.
 \end{equation}
Equation \eqref{Kands} follows from the fact that, for the values of $s$ under consideration $\floor{N/K} = \floor{N/s}$. From \eqref{Kands}, we can say,
 \begin{equation*}
 	s \geq \frac{N}{\floor{N/K}}-K.
\end{equation*}
Now, we have seen that for $s \geq \frac{N}{\floor{N/K}}-K$, Equation \eqref{NgeqK} holds. If $\frac{N}{\floor{N/K}}-K \leq \floor[\Big]{\frac{N}{\ceil{N/K}}}+1$, the we can say that \eqref{NgeqK} holds for all $s\in [1:K]$. Define $s_{min} \triangleq \frac{N}{\floor{N/K}}-K$ and $s_{max} \triangleq \floor[\Big]{\frac{N}{\ceil{N/K}}}+1$.

Define $\gamma \triangleq \frac{N}{K} \in \mathbb{R}$. When $N>K$, $\gamma>1$. Then, we have,
\begin{subequations}
	\begin{align*}
	s_{min} &= \frac{N}{\floor{\gamma}}-\frac{N}{\gamma}=N\left(\frac{1}{\floor{\gamma}}-\frac{1}{\gamma}\right)\leq \frac{N}{\gamma\floor{\gamma}}.
	\end{align*}
\end{subequations}
Similarly, 
\begin{equation*}
s_{max} = \floor[\Big]{\frac{N}{\ceil{\gamma}}}+1\geq \frac{N}{\ceil{\gamma}}.
\end{equation*}
When $\gamma \geq 2$, $s_{max}\geq s_{min}$, since $\gamma \floor{\gamma}\geq \ceil{\gamma}$. 

For $\gamma \in (1,2)$, we have,
\begin{subequations}
	\begin{align*}
	s_{min} &= N-\frac{N}{\gamma}\leq \frac{N}{2} \leq \floor[\Big]{\frac{N}{2}}+1 =s_{max}.
	\end{align*}
\end{subequations}
That is, for every $\gamma>1$, $s_{min}\leq s_{max}$.

We have shown that when $N>K$, for all $s\in [1:K]$, $R_{new}^s(M)\geq R_{cut-set}^s(M)$. That means, when $N>K$, $R_{new}(M)\geq R_{cut-set}(M)$.

This completes the proof of Lemma \ref{better}.\hfill $\blacksquare$

\section{Optimal rate-memory trade-off of the $(K=3,L=2,N=3)$ MACC scheme}
\label{App3}
In this section, we provide an achievability result for the $(K=3,L=2,N=3)$ MACC scheme at $M=\frac{2}{3}$. Also, we show that $R^*(\frac{2}{3}) = 1$. 

In the placement phase, the server divides each file into 3 non-overlapping subfiles of equal size. Then, $W_n = \{W_{n,1},W_{n,2},W_{n,3}\}$ for $n\in[3]$. The cache placement is coded, and is as follows:
\begin{align*}
	Z_1 &= \{W_{1,1}\oplus W_{2,1},W_{2,1}\oplus W_{3,1}\}\\
    Z_2 &= \{W_{1,2}\oplus W_{2,2},W_{2,2}\oplus W_{3,2}\}\\
    Z_3 &= \{W_{1,3}\oplus W_{2,3},W_{2,3}\oplus W_{3,3}\}.
\end{align*}

Let $\mathbf{d} = (d_1,d_2,d_3)$ be the demand vector. Then the server transmits $W_{d_1,3},W_{d_2,1}$ and $W_{d_3,2}$. Therefore, the rate of transmission is 1.

The decodability of the files is straight forward. Consider user $\mathcal{U}_1$ demanding $W_{d_1}$. The user directly receives $W_{d_1,3}$ from the server transmission. Since, subfile $W_{d_2,1}$ is available from the transmission, user $\mathcal{U}_1$ can get $W_{n,1}$ for all $n\in [3]$ using $Z_1$, which includes $W_{d_1,1}$. Using $W_{d_3,2}$ and $Z_2$, user $\mathcal{U}_1$ can decode $W_{d_1,2}$. Similarly, all the users can get their demanded file without any uncertainty. Therefore, the memory-rate pair $(\frac{2}{3},1)$ is achievable.

For the $(K=3,L=2,N=3)$ MACC scheme, the memory-rate pairs $(0,3)$ (trivial) and $(\frac{3}{2},0)$ (see \cite{NaR}) are achievable. By using the scheme in \cite{SPE}, the memory-rate pair $(1,\frac{1}{3})$ is achievable. Therefore, by memory sharing $R(M)= 3(1-M)$ for $0\leq M\leq \frac{2}{3}$, $R(M) = \frac{7}{3}-2M$ for $\frac{2}{3}\leq M\leq 1$ and $R(M) =  1-\frac{2M}{3}$ for $1\leq M\leq \frac{3}{2}$ are achievable. In Example \ref{example2}, we showed that, $R^*(M)\geq 3(1-M)$, $R^*(M)\geq \frac{7}{3}-2M$ and $R^*(M)\geq  1-\frac{2M}{3}$. Therefore, we have,
\begin{equation*}
	R^*(M) = 
	\begin{cases}
		3(1-M) &\text{ for } 0\leq M\leq \frac{2}{3}.\\
		\frac{7}{3}-2M &\text{ for } \frac{2}{3}\leq M\leq 1.\\
		1-\frac{2M}{3} &\text{ for } 1\leq M\leq \frac{3}{2}.
	\end{cases}
\end{equation*}

\end{appendices}

\section*{Acknowledgment}
This work was supported partly by the Science and Engineering Research Board (SERB) of Department of Science and Technology (DST), Government of India, through J.C. Bose National Fellowship to B. Sundar Rajan.

\end{document}